\documentclass[aip,jap,aps,prb,twocolumn,reprint,floatfix,showkeys,superscriptaddress,showpacs,nofootinbib]{revtex4-1}
\pdfoutput=1

\usepackage[english]{babel}
\usepackage[utf8]{inputenc}
\usepackage{graphicx}
\usepackage{subfigure}
\usepackage{filecontents}
\usepackage{amsmath,ragged2e}
\usepackage{amsmath,ragged2e}
\usepackage{amssymb}
\usepackage{bm}
\usepackage{color}
\usepackage[normalem]{ulem}

\usepackage{latexsym}
\usepackage{dcolumn}
\usepackage{amsxtra}

\newcommand{\bn}[1]{\mbox{\boldmath $#1$}}

\def\mb{\mbox}
\def\e{\mathop{\rm \mbox{{\Large e}}}\nolimits}

\def\mb{\mbox}

\begin{document}

\preprint{}

\title{A new phenomenon in graphene: The pseudospinorial Zitterbewegung.}

\author{L. Diago-Cisneros}
\email[E-mail and orcid:$\;$]{ldiago@fisica.uh.cu \& leovildo.diago@ibero.mx \& http/orcid.org/0000-0001-9409-1545}
\affiliation{Facultad de F\'{\i}sica, Universidad de La Habana, Cuba}
\affiliation{Departamento de F\'{\i}sica y Matem\'{\i}ticas, Universidad Iberoamericana, CDMX, M\'{e}xico}
\author{Eduardo Serna}
\email{sernaed95@gmail.com}
\affiliation{Departamento\& de F\'{\i}sica y Matem\'{a}ticas, Universidad Iberoamericana, CDMX, M\'{e}xico}
\author{I. Rodr\'{\i}guez Vargas}
\email[E-mail and orcid:$\;$]{isaac@uaz.edu.mx \& http/orcid.org/0000-0003-0087-8991/}
\affiliation{Unidad Acad\'{e}mica de F\'{\i}sica, Universidad Aut\'{o}noma de Zacatecas, Zac., M\'{e}xico}
\author{R. P\'{e}rez-\'{A}lvarez}
\email[E-mail and orcid:$\;$]{rpa@uaem.mx \& http/orcid.org/0000-0003-1119-1159/}
\affiliation{Centro de Investigaci\'{o}n en Ciencias, Universidad Aut\'{o}noma del Estado de Morelos, Cuernavaca, M\'{e}xico}

\begin{abstract}
We foretell a new pseudospin-dependent phenomenon in mono-layer graphene (MLG), which is numerically simulated \emph{via} an innovator nano-spintronic device. We proposed a novel theoretical procedure for describing the dynamics of Dirac fermions, departing from classic theoretical modelling. More importantly, we have found appealing evidences of wiggling anti-phase oscillations in the probability density time-distribution for each sub-lattice state, which we called pseudospinorial Zitterbewegung effect (PZBE). The PZBE undergoes modulated by a robust transient character, with decay time of femtoseconds. Interestingly, several features of the PZBE become tunable, even up to fully vanishing it at the vicinity of the Dirac points, as well as for a symmetric pseudospin configuration. We have observed evidences of perfect Klein tunneling and perfect anti-Klein backscattering in a single simulation, which is unprecedented for Q1D-MLG, as far as we know.
\end{abstract}


\maketitle

\section{\label{sec:Intro} Introduction}

On the last decades, graphene has shown to be a material with very useful properties for spintronics. Intensive studies were addressed to this topic, looking for the miniaturization, as well as for the improvement on the efficiency of graphene-based devices; taking advantages of its low-energy excitation\cite{Richter2012} and a better energetic exploitation\cite{Fabian2014}.  Another intriguing characteristics of the graphene are the chirality, the gapless spectrum and spin-charge carriers that \textit{mimic} massless relativistic particles called Dirac's fermions\cite{Geimk2007,Castro2009,Vasilievna2008}.

\begin{figure}[ht!]
  \includegraphics[width=\linewidth]{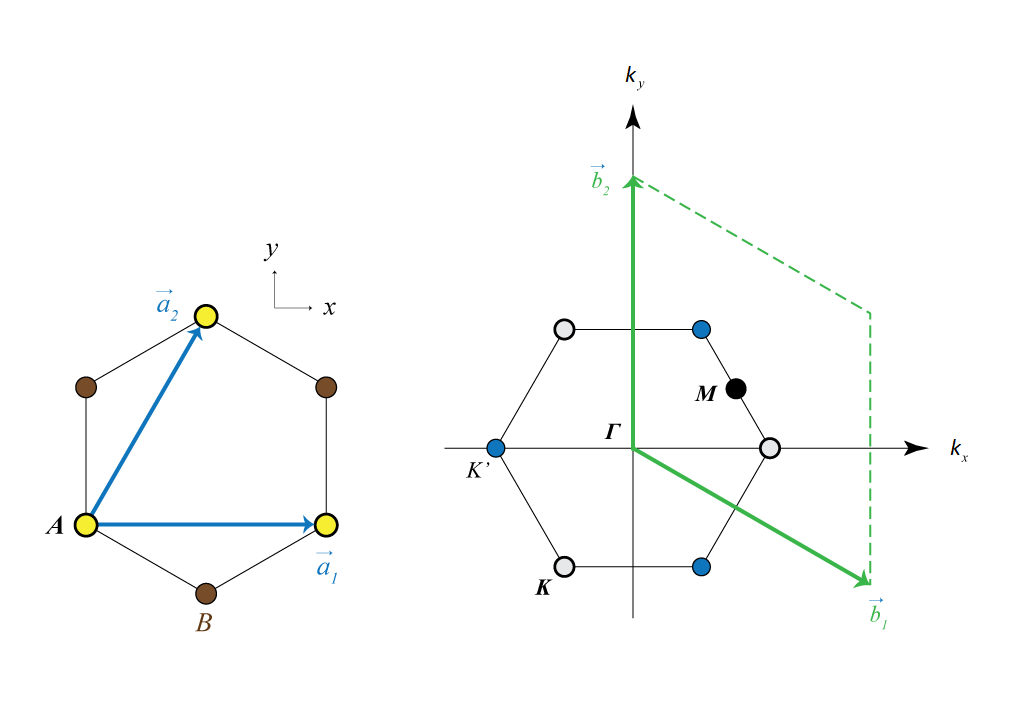}
  \caption {\label{fig:cells} (Color online) Mono-layer graphene lattice and its Brillouin Zone. Left panel: Displays the sub-lattice sites for atoms $A$ and $B$, whereas $\vec{a}_1$ and $\vec{a}_2$ are the lattice unit vectors. Right panel: Shows a standardized view for the Brillouin Zone. The Dirac's points $K(K^{'})$ allocation are refered to the high-symmetry $\Gamma$ point, whose coordinates are $k_{x} = \pm(4\pi/3a_{0})$ and $k_{y} = 0$.}
\end{figure}

Since the early elucidation by Schr\"{o}dinger\cite{Schro1930}, back in the age of the quantum mechanics rise, the free-space relativistic-electron \textit{trembling} motion, is still one of the relevant anomalous phenomena present in solid materials and being somewhat puzzling, remains yet cryptic to detect experimentally. According to Schr\"{o}dinger's ideas, the relativistic-electron position experiences rapid periodic oscillations, which he named after \textit{Zitterbewegung} effect (ZBE). However, years later several aspects were reexamined by Dirac\cite{Dirac1958} and the average quantities over the Compton scale had to be invoked to recover meaningful results\cite{Frolova2008,Biswas2014}. On the contrary, it have been demonstrated that relativistic-electron spatial localization can be narrower than the Compton wavelength, together with the lack of any ZBE because the pair $e^{-}-e^{+}$ interaction is forbidden by the relativistic quantum-field theory\cite{Krekora2004}. These outstanding pioneering works\cite{Schro1930,Dirac1958}, have inspired a large number of theoretical studies in different physical systems, for instance: in semiconductors\cite{Katsnelson2006,Sidhart2008}$^{\footnotesize \mb{[and references therein]}}$, and in graphene\cite{Frolova2008,Rusin2008,Rusin2011,Richter2012}; pursuing a deeper insight into the controversial ZBE.

The Rashba spin-orbit interaction (SOI-R) role for the development of spintronics in graphene, is a tremendous challenge, because the SOI-R is quite imperceptible in that carbon allotrope\cite{Geimk2007}. Meanwhile, this question has been also reviewed in other studies\cite{MacDonald2006,Fabian2014}, which claimed that the SOI-R makes the graphene a very good candidate to be used in the management of \textit{qbits}. Besides, there is a belief that could exist an interplay between the ZBE and the SOI-R\cite{Biswas2014}, following the sensible dependence of the ZBE from the SOI-R strength in $III-V$ semiconducting quantum wells.

\begin{figure}[ht!]
\includegraphics[width=\linewidth]{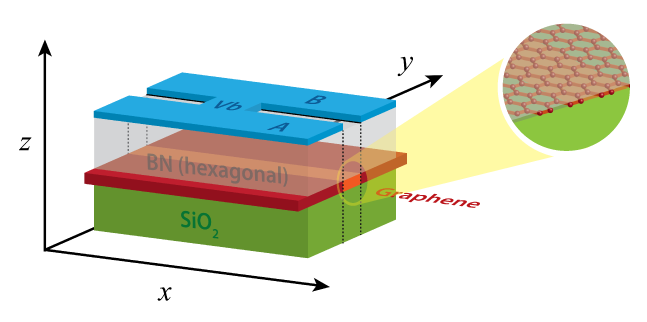}
 \caption{\label{fig:modfis} (Color online) Graphene system (red), with a lithographically printed nanoribbon (middle yellow strip) located between two materials that do not affect the charge carrier's dynamics, in this case: Hexagonal Boron Nitride ($hBN$, transparent gray) and Silicon Dioxide ($SiO_{2}$, green). The charge carriers propagate along the $x$ direction. The movement is limited to one dimension thanks to the charged plates $A$ and $B$ (blue). While the electrically charged plate $V_b$ (blue) represents the potential barrier, where the SOI-R coupling is included.}
\end{figure}

We get motivated by these open topics. In this context, our theoretical approach of the dynamics for Dirac-fermions (DF) in a mono-layer graphene nanoribbon, could contribute to a better understanding of the pseudospinorial Zitterbewegung effect (PZBE), described here. This is the main purpose of the present study. As a collateral goal, we would like to focus the question whether or not the SOI-R imposes any trace over the envisioned under-barrier scattering of DF and/or onto the free-space PZBE as well.

The remaining part of this paper is organized as follows: Section \ref{sec:PhysMod} presents the physical model under study. Section \ref{sec:TeoMod} explains briefly the theoretical approach and the mathematical tools to solve it. Further numerical results are discussed in Section \ref{sec:DisRes}. Finally, in Section \ref{sec:Sum} we sum up.

\section{\label{sec:PhysMod} Physical Model}

The mono-layer graphene (MLG) is composed by a semi-flat single-carbon-atom sheet, with a honeycomb-like lattice [see Fig. \ref{fig:cells}]. The specific MLG structure, which fairly reproduces the envisioned system of the present study, is depicted in Fig. \ref{fig:cells}, and can be described by a triangular lattice with a basis of two atoms \emph{per} unit cell. Accordingly, the unit-cell vectors are $\vec{a}_{1} = a_{0}(1,0)$ and $\vec{a}_{2} = a_{0}/2(1,\sqrt{3})$\cite{Castro2009}, which generate the unit-cell vectors of the Brillouin Zone (BZ), that are given by $\vec{b}_1=4\pi/(2a_{0}\sqrt{3})(\sqrt{3},-1)$ and $\vec{b}_{2} = 4\pi/(a_{0}\sqrt{3})(0,1)$, with $a_{0} = \sqrt{3}a$ where $a\approx 1.42$ \AA,\space is the carbon-carbon distance. From the BZ represented in Fig. \ref{fig:cells}, we can find the Dirac's points allocated at the following ($k_{x},k_{y}$) pair coordinates of the reciprocal laticce: $K = (0,0)$; and $K^{'} = \pm\left(8\pi/3a_{0},0\right)$.

\begin{figure}
 \includegraphics[width=\linewidth]{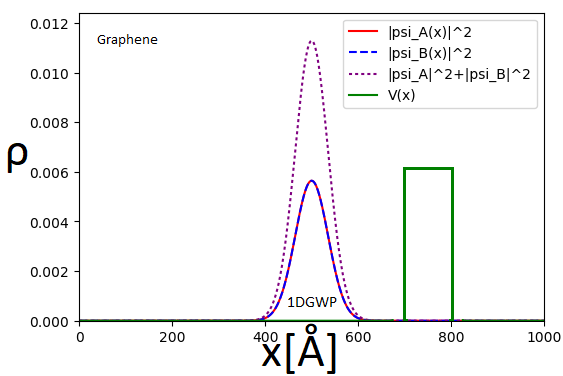}
 \caption{\label{fig:gwp} System in its initial state $t=0$ fs, where the green line represents the potential in the system ($V\neq 0$ eV only in the interval $x=[700,800]$ \AA\space and $V=0$ eV on everything else). The red line is the probability density of the A component, the blue one is for the B component and the purple one is the sum of both components of the pseudo-spinor.}
\end{figure}

The physical system that will be studied in the present work is shown in Fig.  \ref{fig:modfis}. It is intended to have a quasi-one dimensional (Q1D) quantum nanoribbon of MLG, which would be the middle strip (yellow), previously patterned in a two-dimensional sheet (red), placed between $SiO_2$ (green) and hexagonal $hBN$ (transparent gray). We consider biased plates $A$ and $B$ (blue) on top, that will lithographically constrain the DF in the Q1D quantum nanoribbon. In addition, we defined an arbitrary square-potential barrier located at three quarters of the Q1D quantum nanoribbon, represented by a third top-plate $V_{b}$ (blue), that will sectionally modulate the SOI-R strength over the DF, within the under-barrier region. This way, we are able to explore the dynamics of Dirac-fermions betoken as a one-dimensional Gaussian wave-packet (1DGWP) drifting in the free-space region, as well as inside the scatterer, with a group velocity of two order less than light speed. The Fig. \ref{fig:gwp} sketches an instant of this phenomenology.

\section{\label{sec:TeoMod} Theoretical and Mathematical Outlines}

We have assumed free massless DF at low energies. Then, near the $K (K')$ points, the MLG Dirac-like Hamiltonian becomes

\begin{equation}
 \label{eq:grafeno}
 \hat{\bn H}_{G} = v_{\mb{\tiny F}}\vec{\mathbf{\sigma}}\cdot\vec{p},
\end{equation}

\noindent with $\vec{\mathbf{\sigma}} = \hat{\mathbf{\sigma}}_{x}\hat{i} + \hat{\mathbf{\sigma}}_{y}\hat{j}$, being the Pauli matrices $\hat{\mathbf{\sigma}}_{x} = \bigl(\begin{smallmatrix}0&1 \\ 1&0\end{smallmatrix} \bigr)$, $\hat{\sigma}_{y} = \bigl(\begin{smallmatrix}0&-i \\ i&0\end{smallmatrix} \bigr)$, and $\vec{p}=\hat{p}_{x}\hat{i}+\hat{p}_{y}\hat{j}$, is the momentum operator, whose $x, y$ components read $\hat{p}_{x} = -i\hbar\frac{\partial}{\partial x}$ and $\hat{p}_{y} = -i\hbar\frac{\partial}{\partial y}$, respectively. Hereinafter $v_{\mb{\tiny F}}$ stands for the Fermi velocity of the carriers in MLG. Finally, the dispersion law has the widely-known form $E = \pm|\hbar k|v_{\mb{\tiny F}}$. Another striking characteristic of the MLG, is the well-known chirality or helicity. Indeed, the momentum operator for the DF in the sub-lattice \textit{A}, has a preferential settling of its projection along the pseudo-spin direction. While for the DF in the sub-lattice \textit{B}, it is found completely opposite. For this reason the  pseudo-spinor is a two-component state of the form

\begin{equation}
 \label{eq:pseudoespinor}
 \mathbf{\Psi}(x,t) = \binom{\psi_{A}(x,t)}{\psi_{B}(x,t)}.
\end{equation}

\subsection{\label{subsec:SOI} Rashba Spin-Orbit Interaction}

One of the remarkable contributions to the spin manipulation without an external magnetic field, is the SOI-R effect. On general grounds, we depart from a Rashba Hamiltonian for heavy holes\cite{RDiago2010}

\begin{equation}
 \label{eq:SOIR-1}
 \hat{\bn H}_{\mb{\tiny SOI-R}} = -i\alpha\left[\frac{\hat{k}_{-}^{2}}{\hat{k}_{+}}\sigma_{+} -\frac{\hat{k}_+^2}{\hat{k}_{-}}\sigma_{-}\right] - i\beta\left[\hat{k}_{-}^{3}\sigma_{+} - \hat{k}_{+}^{3}\sigma_{-}\right],
\end{equation}
\noindent where $\hat{k}_\pm=(\hat{k}_x\pm i\hat{k}_y)$, $\sigma_\pm=\frac{1}{2}(\sigma_x\pm i\sigma_y)$, $\alpha$ is the linear Rashba parameter and $\beta$ is the cubic Rashba parameter (both of them are material-dependent constants). After some algebra, and by zeroing the components accompanying $\hat{k}_{y}$, we obtain \cite{RCDiagoEPL2015}

\begin{equation}
 \label{eq:SOIR}
 \hat{\bn H}_{\mb{\tiny SOI-R}}=(-i\alpha\partial_{x} + i\beta\partial^{3}_{x})\sigma_{y},
\end{equation}
\noindent being $\partial_{x} = \frac{\partial}{\partial x}$. Thus, the Eq. \eqref{eq:SOIR} is used to address the above posted question, about the SOI-R influence over the envisioned under-barrier scattering of DF, together with the interplay it could possible exert on the PZBE.

\subsection{\label{subsec:FinDif} Finite Differences Scheme}

The Hamiltonian to be used for facing-off the unsettled topics posted in Section \ref{sec:Intro}, in the system depicted by Fig. \ref{fig:modfis}, will be the summation of the Dirac-like Hamiltonian \eqref{eq:grafeno}, with the components $\hat{k}_{y} = 0$, the SOI-R Hamiltonian \eqref{eq:SOIR} and the potential $V(x)$ that represents our quantum barrier (QB), \textit{i.e.},

\begin{widetext}
 \begin{equation}
  \label{eq:hammat}
  \hat{\bn H} = \hat{\bn H}_{G} + \hat{\bn H}_{\mb{\tiny SOI-R}} + V(x)\mathbb{I}_{2} = \begin{pmatrix}V(x)&-i\hbar v_f\partial_x-\alpha\partial_x+\beta\partial^3_x\\-i\hbar v_f\partial_x+\alpha\partial_x-\beta\partial^3_x & V(x)\end{pmatrix},
\end{equation}
\end{widetext}
\noindent where $\mathbb{I}_{2}$ stands for the second order identity matrix. We proceed further discussing the time-dependent Dirac-like equation in a Q1D space, though written it as an evolution equation within the standardized Schrödinger frame.  The time evolution in the free(scattering)-space region, resulting from the diffusion under arbitrary initial conditions, can be an excellent workbench to the finite difference scheme. So, for the mathematical procedure, we dicretize the Q1D Dirac-like equation in the framework of the familiar Schrödinger model

\begin{equation}
 \label{eq:SchDifin}
 (\hat{\bn H}\Psi(x,t))_{j}^{n} = \imath\hbar\frac{\partial}{\partial t}\mathbf{\Psi}(x,t)_{j}^{n},
\end{equation}

\noindent where $n$ is the discretization variable of time, while $j$ is the discretization variable of space, so that $t\rightarrow n\delta t$, and $x\rightarrow j\delta x$; for $n = 1,2,...,N$, and $j = 1,2,...,J$, respectively. The elemental quantity $\delta x (\delta t)$ quotes the space(time) minimal step of the grid. The general solution for \eqref{eq:SchDifin}, can be proposed in the form

\begin{equation}
 \label{eq:tempevo}
  {\bn \Psi}(x,t) = \hat{\bn U}(t,t_{0}){\bn \Psi}(x,t_{0}),
\end{equation}
\noindent where $\hat{\bn U}(t,t_{0})$ is the time-evolution operator, which is defined as

\begin{equation}
 \label{eq:TimOpe}
 \hat{\bn U}(t,t_{0}) = \e^{\frac{i}{\hbar}\delta t\hat{\bn H}}.
\end{equation}

First, it is convenient to use the Cayley's approach\cite{RDiago2010}
\begin{equation}
 \label{eq:Cay}
 \hat{\bn U}(t,t_{0}) = \frac{2\mathbb{I}_{2}}{\mathbb{I}_{2} + \frac{1}{2}\frac{\imath\delta t}{\hbar}\hat{\bn H}} - \mathbb{I}_{2},
\end{equation}
\noindent and next we substitute in \eqref{eq:tempevo}, then we obtain

\begin{equation}
 \label{eq:evotemp1}
 {\bn \Psi}(x,t) = {\bn \Phi}(x,t_{0}) - {\bn \Psi}(x,t_{0}),
\end{equation}

\begin{equation}
 \label{eq:LALO1}
 \mb{where}\;\;\; {\bn \Phi}(x,t_{0}) = \frac{2{\bn \Psi}(x,t_{0})}{\mathbb{I}_{2} + \frac{1}{2}\frac{\imath\delta t}{\hbar}\hat{\bn H}}.
\end{equation}

It is straightforward from \eqref{eq:LALO1}, that
\begin{equation}
 \label{eq:inicial}
 2{\bn \Psi}(x,t_{0}) = \bn{\Phi}(x,t_{0}) + \frac{\imath\delta t}{2\hbar}\hat{\bn H}\bn {\Phi}(x,t_{0}),
\end{equation}
\noindent which is solved for $\bn{\Phi}(x,t_{0})$, to be further substituted in \eqref{eq:evotemp1}, whose discrete form can be cast as

\begin{equation}
\label{eq:evotemp1dis}
\mathbf{\Psi}_j^{n+1}=\mathbf{\Phi}_{j}^n-\mathbf{\Psi}_j^{n}.
\end{equation}

Now we impose the initial boundary conditions

\begin{equation}
 \label{for:inicon}
 \bn{\Psi}_{j=0}^{n=0} = \bn{\Psi}_{j=J}^{n=0}=0,
\end{equation}
\noindent to calculate the wave functions along the Q1D nanoribbon, sketched in Fig. \ref{fig:modfis}(yellow strip). The standard Taylor development is performed for functions evaluated in $j-2$; $j-1$; $j+1$ and $j+2$. Afterward, we solve the equation system for the derivatives and we get

\begin{align}
  \psi_j'&=\frac{\psi_{j-2}-8\psi_{j-1}+8\psi_{j+1}-\psi_{j+2}}{12\delta x}, \label{eq:taylor1}\\
  \psi_j''&=\frac{-\psi_{j-2}+16\psi_{j-1}-30\psi_{j}+16\psi_{j+1}-\psi_{j+2}}{12\delta x^2},
  \label{eq:taylor2}\\
  \psi_j'''&=\frac{-\psi_{j-2}+2\psi_{j-1}-2\psi_{j+1}+\psi_{j+2}}{2\delta x^3},
  \label{eq:taylor3}\\
  \psi_j''''&=\frac{\psi_{j-2}-4\psi_{j-1}+6\psi_{j}-4\psi_{j+1}+\psi_{j+2}}{\delta x^4}.
  \label{eq:taylor4}
\end{align}

By substituting \eqref{eq:hammat} in \eqref{eq:inicial}, using \eqref{eq:taylor1}-\eqref{eq:taylor4}; after adding and subtracting both components of \eqref{eq:inicial}, we finally reach the crucial equation systems

\begin{widetext}
 \begin{align}
  \label{eq:sist1}
  2(\psi_{Aj} + \psi_{Bj}) & = M A_{b} P_{j-2} - 8M A_{b} P_{j-1} + (1+M V)P_{j} + 8M A_{b} P_{j+1} - M A_{b} P_{j+2}\\
  & + M B_{a} Q_{j-2}-M C_{a} Q_{j-1} + M C_{a} Q_{j+1} - M B_{a} Q_{j+2}, \nonumber\\
  \label{eq:sist2}
  2(\psi_{Aj} - \psi_{Bj}) & = -M B_{a} P_{j-2} + M C_{a} P_{j-1} - M C_{a} P_{j+1} + M B_{a} P_{j+2}\\
  & - M A_{b} Q_{j-2} + 8M A_{b} Q_{j-1} + (1+M V)Q_{j} - 8M A_{b} Q_{j+1} + M A_{b} Q_{j+2},\nonumber
 \end{align}
\end{widetext}
\noindent where $M = (\imath\delta t)/(2\hbar)$, $A_{b} = (A_a)/(12\delta x)$, $A_{a} = -\imath\hbar v_{f}$, $B_{a} = (\alpha\delta^{2}x + 6\beta)/(12\delta^{3}x)$, and $C_{a} = (2\alpha\delta^{2}x + 3\beta)/(3\delta^{3}x)$. The quantities we are looking for, are $P_{j}$ and $Q_{j}$, since
\begin{equation}
 \label{for:phiAB}
  \phi_{(A,B)j} = \frac{1}{2}\left[P_{j} \pm Q_{j}\right].
\end{equation}
\noindent Considering that we have $P_{j}$ and $Q_j$ in both equation systems \eqref{eq:sist1} and \eqref{eq:sist2}, they can be treated as $(J\times 1)$ vectors, as well as the terms $2(\psi_{Aj} + \psi_{Bj})$ and $2(\psi_{Aj} - \psi_{Bj})$. Besides, for convenience, the physical-system's constants will be embedded in the $(J\times J)$ matrices $\mathbb{M}_{A,B}$; which are $5$-diagonal ones. Thus, we define

\begin{align}
 \label{eq:sist3}
 \bn{\Psi}_{+} & = \mathbb{M}_{A}\cdot \bn{P} + \mathbb{M}_{B}\cdot \bn{Q}\\
 \label{eq:sist4}
 \bn{\Psi}_{-} & = \mathbb{M}_{B}^{T}\cdot \bn{P} + \mathbb{M}_{A}^{T}\cdot \bn{Q}.
\end{align}
\noindent and after not too hard algebra, one can obtain

\begin{widetext}
\begin{align}
 \label{eq:sist5}
  \bn{Q} & = \left[\mathbb{M}_{A}^{T} - \mathbb{M}_{B}^{T}\cdot \left(\mathbb{M}_{A}^{-1}\cdot \mathbb{M}_{B}\right)\right]^{-1}\cdot\left[\bn{\Psi}_{-} - \mathbb{M}_{B}^{T}\cdot \left(\mathbb{M}_{A}^{-1}\cdot \bn{\Psi}_{+}\right)\right], \\
 \label{eq:sist6}
  \bn{P} & = \mathbb{M}_{A}^{-1}\cdot\left(\bn{\Psi}_{+} - \mathbb{M}_{B}\cdot\bn{Q}\right).
\end{align}
\end{widetext}

In short, to simulate the time evolution, resulting from the quantum diffusion(scattering) of DF, whenever they move in the free(QB) regions of a MLG, we first solve \eqref{eq:sist5} and \eqref{eq:sist6} for $P_j$ and $Q_j$, respectively, next we find $\phi_{Aj}$ and $\phi_{Bj}$ from \eqref{for:phiAB}, to substitute afterward in \eqref{eq:evotemp1dis}. The later procedure is a cyclic $(n = 1,\ldots N)$-time loop for each of the $(j = 1,\ldots J)$-space points.

\section{\label{sec:DisRes} Discussion of Results}

We have gathered wordline movies of the envisioned DF quantum diffusion(scattering), for several physical conditions of interest, the correspondent multimedia are available to be consulted and/or downloaded from a permanent web link [see Ref.\onlinecite{AnimaQuantrix2018}]. For the timeline animations we used a box length of $L = 500$ \AA [see the abscise axis of Fig-\ref{fig:gwp}], and a simulation time of $50$ fs. The centroid of the Gaussian wave-packet is allocated at $x_{0} = 250$ \AA, while the dispersion size was fixed as $\sigma = 50$\AA, and a barrier thickness of $\Delta x_{b} = 100$ \AA$\,$ had been taken. It is important to mention that, for the quantum diffusion (scattering) to be stable, we must follow the requirement\cite{Carrillo2015}

\begin{equation}
 \label{for:reqXT}
  \delta t\leq \frac{\delta^{2} x}{2},
\end{equation}
\noindent which as a bonus becomes unexpectedly relevant in our study, provided we have demonstrated its influence to resolve several fine features of the wave-packet diffusion to compare with\cite{Frolova2008}.
For describing the dynamics of the DF in a MLG-based Q1D nanoribbon [see Fig. \ref{fig:modfis}], we have represented them by a 1DGWP defined as

\begin{equation}
 \label{eq:1DGWP}
  \bn{\Psi}(x,t=0) = \frac{\xi}{\sigma}\sqrt[]{\frac{1}{\pi}}\e^{-\frac{(x-x_{0})^{2}}
  {2\sigma^{2}}}\e^{\imath k(x-x_{0})},
\end{equation}
\noindent whose group velocity decreases proportional to $\delta t$, remaining shorter than light speed. The following set of the real-spin shape

\begin{center}
\begin{tabular}{c c c c}
 $\xi = \binom{1}{0};\;$ & $\xi = \frac{1}{\sqrt{2}}\binom{1}{1};\;$ & $\xi = \frac{1}{\sqrt{2}}\binom{1}{i};\;$ & and $;\;\xi = \frac{1}{\sqrt{2}}\binom{1}{e^{i\pi/4}}$, \\
\end{tabular}
\end{center}
\noindent unambiguously determines the four different cases of the initial pseudo-spinor configuration (PSC) we are interested, for comparison with a prior study\cite{Frolova2008}, pursuing a validation of our model.

In this section, we divide the discussion in three parts. Firstly, there is a semi-quantitative comparison for a Dirac-$e$ 1DGWP in the free space, between our results and those published elsewhere\cite{Frolova2008}. Secondly, we analyze the dynamics of the 1DGWP during its drifting, when sampling the vicinity of the Dirac points $K$ and $K'$ [see Fig. \ref{fig:cells}], as well as far from them. Finally, we try to study whether the under-barrier biased SOI-R, interplays during the DF interaction with the QB or not.

\begin{figure}
\begin{tabular}{|c|}
      \hline
			\\
			
			\includegraphics[width=0.8\linewidth]{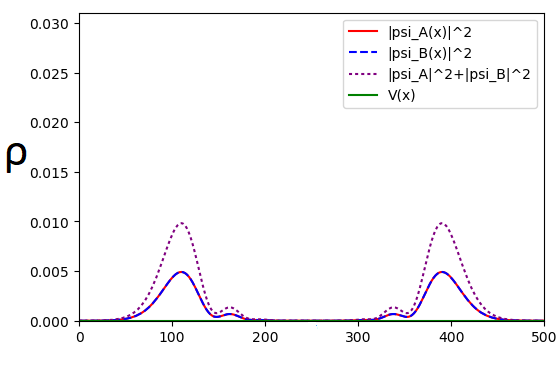} (a)\\
			$ \xi=\binom{1}{0}$ \\

			\includegraphics[width=0.8\linewidth]{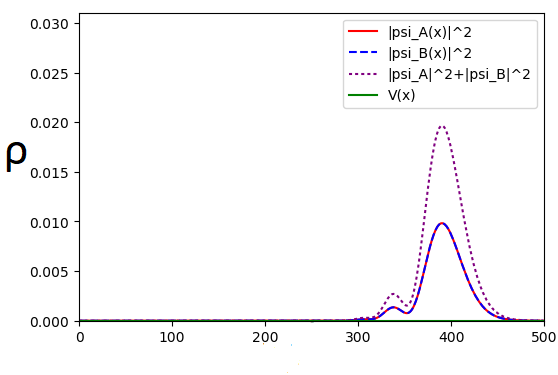} (b)\\
			
            $\xi=\frac{1}{\sqrt{2}}\binom{1}{1}$ \\

			\includegraphics[width=0.8\linewidth]{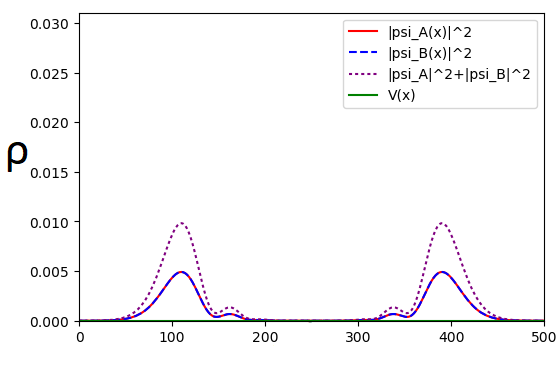} (c)\\
			
            $\xi=\frac{1}{\sqrt{2}}\binom{1}{i}$ \\
			
			\includegraphics[width=0.8\linewidth]{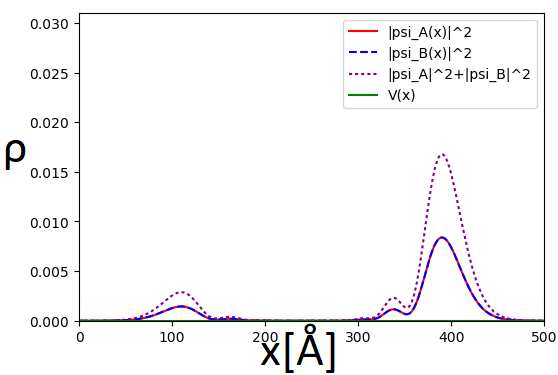} (d)\\
			
            $\xi=\frac{1}{\sqrt{2}}\binom{1}{e^{i\pi/4}}$ \\
      \hline
\end{tabular}
\caption {\label{fig:comp} (Color online) Worldline dynamical drift of our 1DWGP at $t = 20$ fs and given values of $\xi$, for a qualitative comparison with the wave packet at a similar time interval, and with same $\xi$ configurations as in reference [\onlinecite{Frolova2008}].}
\end{figure}

In Figure \ref{fig:comp} we show the timeline evolution in the free-space region, resulting from the quantum diffusion of Dirac-$e$ modeled by the 1DGWP \eqref{eq:1DGWP}, on the surface of a MLG narrow nanoribbon, that we have assumed a width less than the lattice parameter. As can be seen, the 1DGWP spreads and ultimately divides into a couple of variable-shape sub-1DGWP, which drift apart respect to the center of the box, as expected \cite{Frolova2008}. In all panels the probability density of each pseudo-spinor components $\rho_{A,B} = |\psi_{A,B}|^{2}$, is represented with a red-solid(blue-dashed) line, respectively. Meanwhile the conservation law $\rho = |\psi_{A}|^{2} + |\psi_{B}|^{2}$ is depicted with a purple-dotted line. Next, a case-by-case comparison between our data with several graphs obtained by Frolova \emph{et al.}\cite{Frolova2008} for identical PSC set, is presented and we remark the good qualitative agreement achieved. For the sake of accuracy, we first ``cut'' their 3D wave packet (WP) along their $x/d$ axis at $y/d = 0$ (see Fig.1 in Ref.[\onlinecite{Frolova2008}]). Thereby, the panels (a)-(d) show a projection of the 1DGWP \eqref{eq:1DGWP} worldline on that cut plane. In panel (a), with $\xi = \binom{1}{0}$, the 1DGWP \eqref{eq:1DGWP} turns divided in two smaller sub-1DGWP with the same amplitude, that begin moving in opposite directions (see Fig.1 in Ref.[\onlinecite{Frolova2008}]). In panel (b), when $\xi = \frac{1}{\sqrt{2}}\binom{1}{1}$, the 1DGWP becomes divided differently, but it does remains moving in the positive direction of $x$ (see Fig.3 in Ref.[\onlinecite{Frolova2008}]). The panel (c) plots the evolution for the PSC $\xi = \frac{1}{\sqrt{2}}\binom{1}{\imath}$ and the splitted sub-1DGWP are similar in the amplitude of $|\psi_{A,B}|^{2}$, though slightly less than that of the panel (a) (see Fig.5 in Ref.[\onlinecite{Frolova2008}]). Finally, the panel (d) displays the case for PSC $\xi = \frac{1}{\sqrt{2}}\binom{1}{e^{i\pi/4}}$, the WP is divided in two sub-1DGWP, however, left-side one is smaller compared to the one on the right-side (see Fig.7 in Ref.[\onlinecite{Frolova2008}]). It is worthwhile to stress, that inner-side small ripples, were detected solely under the fulfillment of \eqref{for:reqXT}, by fixing a time partition of $c\delta t$ with $c \in [0.5,1]$.
Something important to note here is the following, when the 1DGWP is located at the point $K$ on the reciprocal lattice ($k_{x,y} = 0$) [see Fig.\ref{fig:cells}], no oscillations of $|\psi_{A,B}|^{2}$ were detected. Multimedia of the Dirac-$e$ worldline dynamic in the free space, are  available in a permanent web link [see Ref.\onlinecite{AnimaQuantrix2018}].

\begin{figure}
\begin{tabular}{|c|c|}
      \hline
			&\\
			\includegraphics[width=0.4\linewidth]{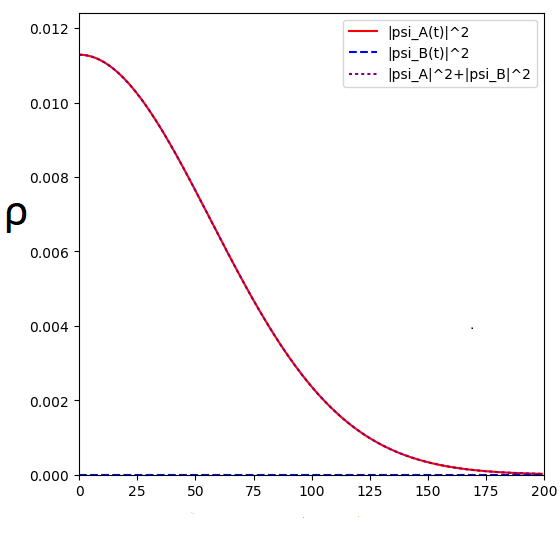} (a)&
			\includegraphics[width=0.4\linewidth]{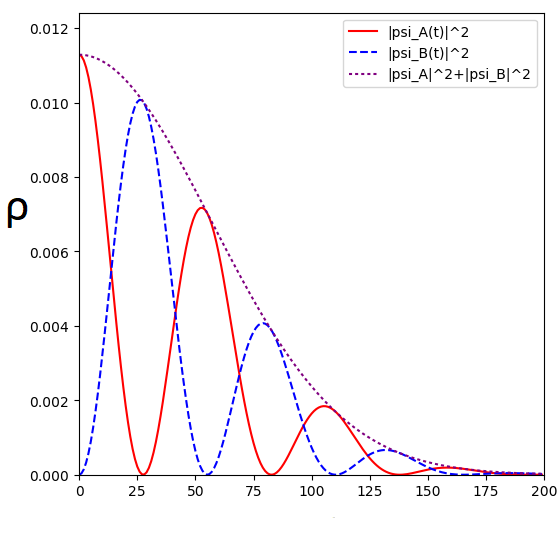} (e)\\
			$\xi=\binom{1}{0}$&$\xi=\binom{1}{0}$\\
      \includegraphics[width=0.4\linewidth]{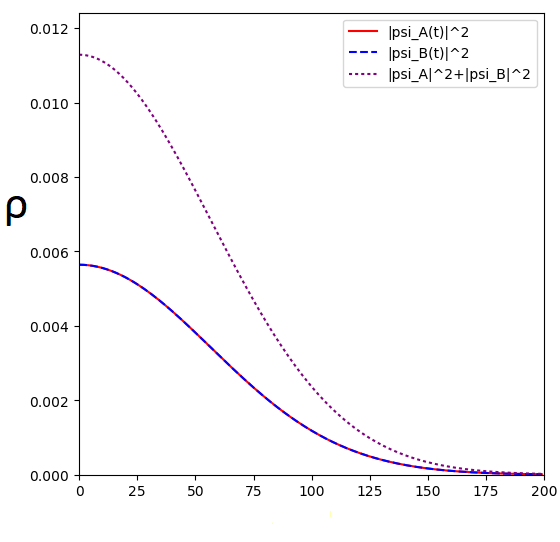} (b)&
			\includegraphics[width=0.4\linewidth]{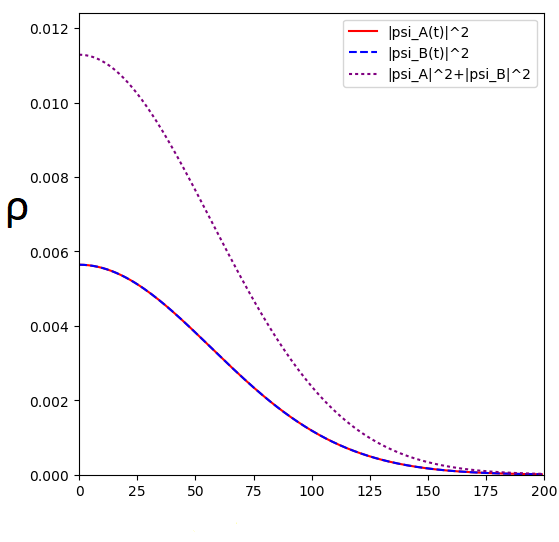} (f)\\
			 $\xi=\frac{1}{\sqrt{2}}\binom{1}{1}$&$\xi=\frac{1}{\sqrt{2}}\binom{1}{1}$\\
      \includegraphics[width=0.4\linewidth]{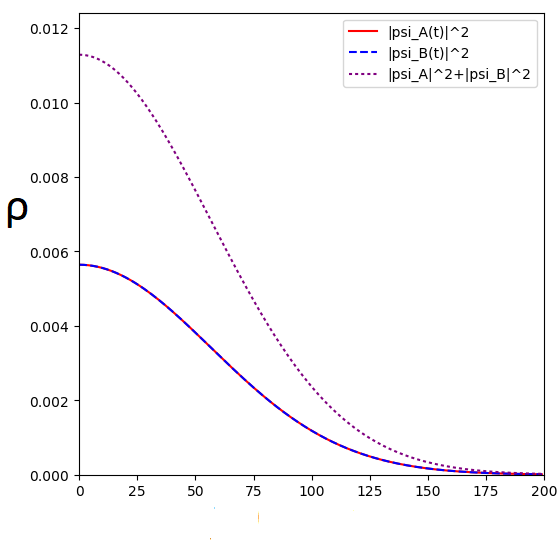} (c)&
			\includegraphics[width=0.4\linewidth]{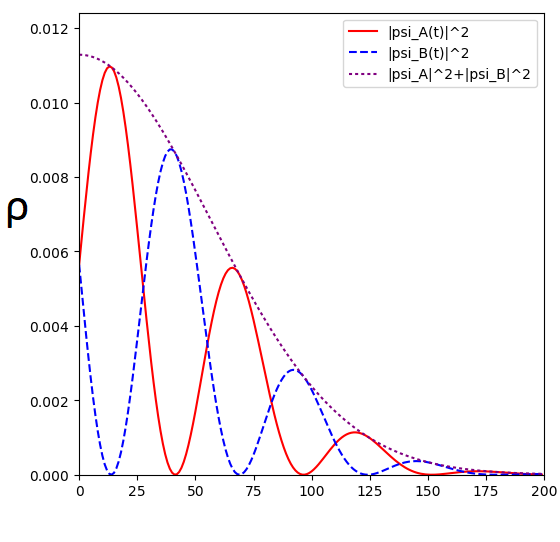} (g)\\
			 $\xi=\frac{1}{\sqrt{2}}\binom{1}{i}$&$\xi=\frac{1}{\sqrt{2}}\binom{1}{i}$\\
			\includegraphics[width=0.4\linewidth]{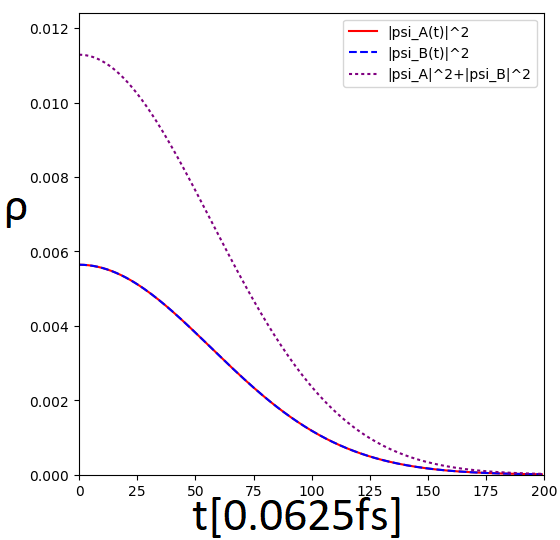} (d)&
			\includegraphics[width=0.4\linewidth]{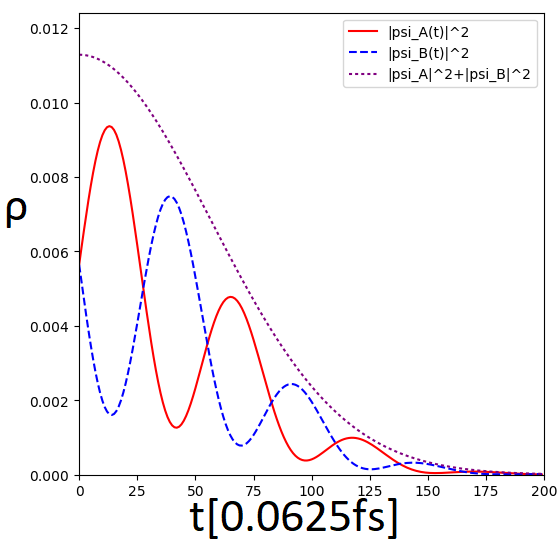} (h)\\
			 $\xi=\frac{1}{\sqrt{2}}\binom{1}{e^{i\pi/4}}$&$\xi=\frac{1}{\sqrt{2}}\binom{1}{e^{i\pi/4}}$\\
      \hline
\end{tabular}
\caption{\label{fig:Zitcomp} (Color online) Qualitative comparison for probability density $|\psi_{A,B}|^{2}$ as function of time, at $k_{x,y} = 0$ (left-column panels), with that of $k_{x(y)} = 0.09(0)$ \AA$^{-1}$ (right-column panels). Panels (e), (f) and (h) show the PZBE.}
\end{figure}

\begin{figure}
 \includegraphics[width=\linewidth]{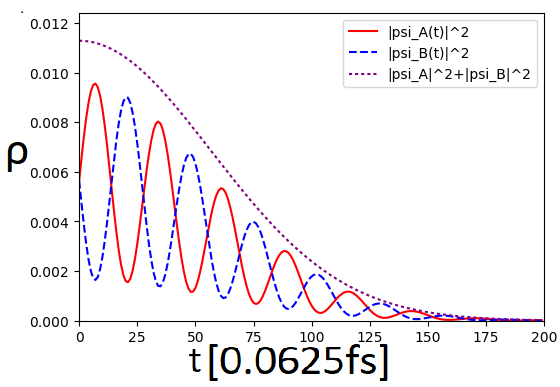}
  \caption{\label{fig:k018} (Color online) PZBE through the evolution of the probability density $|\psi_{A,B}|^{2}$ as a function of time during the first $12.5$ fs of the 1DGWP drift at $k_{x(y)} = 0.18(0)$ \AA$^{-1}$ and $\xi = \frac{1}{\sqrt{2}}\binom{1}{e^{\imath\pi/4}}$}
\end{figure}

Figure \ref{fig:Zitcomp} plots the foremost contribution of the present study, \textit{i.e.}, the stable evidences of anti-phase oscillations in the probability density time-distribution for each pseudo-spinor component $|\psi_{A,B}|^{2}$ of the corresponding sub-lattice state \eqref{eq:pseudoespinor}. These oscillations, resemble those of the wiggling motion discovered by Schr\"{o}dinger\cite{Schro1930}, but in this case for $\rho_{A,B} = |\psi_{A,B}|^{2}$. Thus, we have named the phenomenology described here, as pseudospinorial Zitterbewegung effect [see panels (e), (g) and (h)]. In all panels, we have represented $|\psi_{A}|^{2}$ with a red-solid line and $|\psi_{B}|^{2}$ with a blue-dashed one. For the selected parameters rank, the conservation requirement $\rho = |\psi_{A}|^{2} + |\psi_{B}|^{2}$ was carefully verified and displayed by a purple-dotted line, everywhere. Firstly, the 1DGWP was allocated at the Dirac point $K$ [see Fig.\ref{fig:cells}], afterward it have been released to drift along the $500$ \AA $\;L$-length box described above. No oscillations of $|\psi_{A,B}|^{2}$ were detected at all, despite we explore several PSC [see left-column panels (a)-(d)]. Interestingly, by re-placing now the 1DGWP center mass away from $K$, clear $|\psi_{A,B}|^{2}$ oscillations were detected when it drifts in the box, whose shape strongly depends on the initial PSC, as can be straightforwardly seen from right-column panels: (e), (g) and (h). Indeed, for $\xi = \binom{1}{0}$ [see panel (e)], $|\psi_{A,B}|^{2}$ maximizes(minimizes) at $t = 0$, meanwhile they start from the same amplitude for $\xi = \frac{1}{\sqrt{2}}\binom{1}{\imath}$ [see panel (g)]. Panel (h) shows a similar feature to the later one, though with a reduce amplitude of $|\psi_{A,B}|^{2}$ for $\xi = \frac{1}{\sqrt{2}}\binom{1}{e^{\imath\pi/4}}$.  Having accomplished a semi-empirical method \eqref{eq:rhotempevo1} --to be detailed described below--, and taking all the values of Tab.\ref{tab:nums} into account, the simulation results of right-column panels (e)-(h) were confirmed and accurately described by this methodology. Therefore, since the alternated finite quotations of the constants $D$ and $E$ for some PSC [see Tab.\ref{tab:nums}], the sinusoidal function survives, yielding the PZBE oscillations to rise [see right-column panels (e), (g) and (h)]. During numerical explorations, unexpectedly for a fully symmetric PSC $\xi = \frac{1}{\sqrt{2}}\binom{1}{1}$ (even at $k = 0.09$ \AA$^{-1}$), the PZBE vanishes [see panel (f)]. Again, after the computation with \eqref{eq:rhotempevo1}, this is not surprising given the fact that a simultaneous zeroing of the parameters $D$ and $E$ for $\xi = \frac{1}{\sqrt{2}}\binom{1}{1}$ [see Tab.\ref{tab:nums}], leads the periodic function in \eqref{eq:rhotempevo1} to nullify, and thereby no oscillations can be expected at that PSC. Good qualitative agreement of the PZBE with previous reports of the ZBE in graphene\cite{Frolova2008,Rusin2008,Rusin2011,Richter2012}, has been achieved, because a robust transient character of the oscillations, with decay time of about $10.5$ femtoseconds, was found in all cases under examination here. Multimedia for the trembling dynamics of Dirac-$e$ in the free space, are  available in a permanent web link [see Ref.\onlinecite{AnimaQuantrix2018}] and thus, the rise of the PZBE \emph{via} $|\psi_{A,B}|^{2}$ anti-phase oscillations, can be more explicitly observed in colored worldline movies.

The Figure \ref{fig:k018} is devoted to demonstrate the dependence of PZBE frequency on $k$. Notice the difference in comparison with right-column panels: (e), (g) and (h) of Fig. \ref{fig:Zitcomp}, \emph{i. e.}, the higher $k$ values, the smaller $|\psi_{A,B}|^{2}$ oscillation period. For instance, by letting grow $k$ in about one order, the period diminishes in almost to a half [see panel Fig.\ref{fig:Zitcomp}(h) for purpose of comparison]. Due to the lack of proper explicit analytic relations for such $k$-dependent PZBE oscillations, we have derived through a numerical method, a function that describes accurate enough the oscillating behavior for each $ \rho_{A,B} = |\psi_{A,B}|^{2}$ component, and can be expressed as follow:

\begin{equation}
 \label{eq:rhotempevo1}
  \rho_{s\xi}(500,t) = U_{s}(t)\rho_{s\xi}(500,0),
\end{equation}
\noindent being $U_{s}(t) = \frac{1}{2}\left[\pm C\sin(Dt+E)+1\right]\e^{Ft^2}$, for $s = A,B$; respectively. It is worthwhile stressing, that the procedure \eqref{eq:rhotempevo1} although not an explicit function of $k$, is well-chosen and reliably reproduces the $k$-dependent PZBE oscillating features of Fig.\ref{fig:k018} [see red-solid and blue-dashed lines for each sub-lattice component]. As a bonus, the later relationship can be fairly interpreted as a comprehensive analog with the time-evolution-operator based solution \eqref{eq:tempevo} and in that sense \eqref{eq:rhotempevo1} emerges as a valid model for describing the dynamic properties of the envisioned 1DGWP. By summing up both previous equations, we have

\begin{equation}
 \label{eq:rhotempevo2}
 \rho_{\xi}(500,t) = \e^{Ft^2}\rho_{\xi}(500,0),
\end{equation}
\noindent which is a practical and simplified expression for the envelope probability density $\rho = |\psi_{A}|^{2} + |\psi_{B}|^{2}$ [see purple-dotted line of Fig.\ref{fig:Zitcomp}-Fig.\ref{fig:k018}], preserving though the time-evolution analogy commented above.

\begin{table}[ht]
 \begin{center}
  \begin{tabular}{|c|c c c c|}
  \hline
Constant  & $\xi=\binom{1}{0}$ & $\xi=\frac{1}{\sqrt{2}}\binom{1}{1}$ & $\xi=\frac{1}{\sqrt{2}}\binom{1}{i}$ & $\xi=\frac{1}{\sqrt{2}}\binom{1}{e^{i\pi/4}}$\\ [0.5ex]
\hline
$C$ 					 & $0.00$             & $0.00$             & $0.00$     		 		 & $0.71$\\
\hline
$D$[fs$^{-1}$] & $20.00$            & $0.00$            & $20.00$            & $20.00$\\
\hline
$E$						 & $\pi/2$            & $0.00$             & $0.00$             & $0.00$\\
\hline
$F$[fs$^{-2}$] & $-0.04$            & $-0.04$            & $-0.04$            & $-0.04$\\
\hline
RMSU & $6.27\times10^{-6}$ & $1.24\times10^{-7}$ & $1.91\times10^{-6}$ & $2.20\times10^{-5}$\\
\hline
\end{tabular}
\caption{\label{tab:nums} Numerical estimation for coefficients of \eqref{eq:rhotempevo1} for several initial PSC at $k = 0.09$ \AA$^{-1}$. RMSU stands for the root mean square uncertainty.}
\end{center}
\end{table}

If one releases the 1DGWP at the vicinity of the $\Gamma$ point [see Fig.\ref{fig:cells}], the group velocity diminishes and tends rapidly to zero by approaching it. In other words, the 1DGWP becomes almost static; while the frequency of PZBE oscillations is so high that it would seem at first sight, that there are no oscillations at all [see animations of Ref.\onlinecite{AnimaQuantrix2018}]. We have numerically verified that the parameters $D$ and $F$, are the ones that behave as certain functions of $k$. Indeed, $D$ is directly proportional to $k$, while $F$ shows an inverse proportionality to it. Both of them, straightforwardly define the periodic and the envelope parts, respectively, of time-evolution operator-like functions \eqref{eq:rhotempevo1}-\eqref{eq:rhotempevo2}. Thus, the frequency increment when $k$ diminishes (by approaching to $\Gamma$ point); together with the robust decay of about $10.5$ fs, are very likely because, firstly, $D$ having units of [fs$^{-1}$] belongs to the harmonic function in \eqref{eq:rhotempevo1}; and secondly, the exponential part of \eqref{eq:rhotempevo2} depends on $F$, which is in addition, a negative defined quantity [see Tab. \ref{tab:nums}]. Undoubtedly, a much deeper theoretical analysis is achievable as well as required on this relevant PZBE $k$-dependence. However, we do not believe it would significantly modify our results, and in all likelihood, it would just follow the trends we show here, albeit some refinement could be expected specially in the vicinity of the high-symmetry points of reciprocal space.

Finally, lets turn now to the third part of this section, in which we will try to provide evidences on whether or not the under-barrier biased SOI-R according to \eqref{eq:SOIR}, influences the DF during their tunneling throughout the QB.

\begin{figure}
\begin{tabular}{|c|c|}
      \hline
			&\\
			 \includegraphics[width=0.4\linewidth]{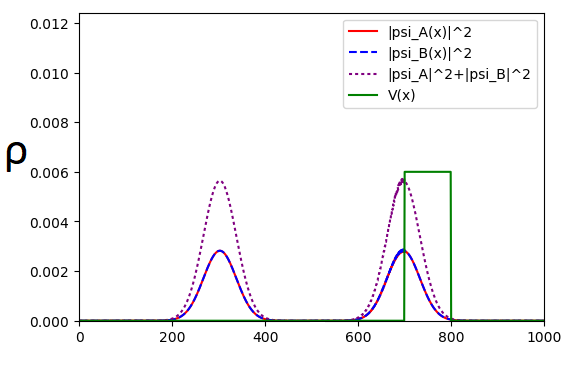}  (a)&
			 \includegraphics[width=0.4\linewidth]{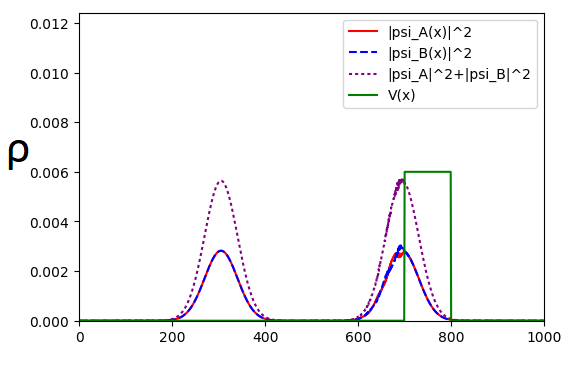} (d)\\
             \includegraphics[width=0.4\linewidth]{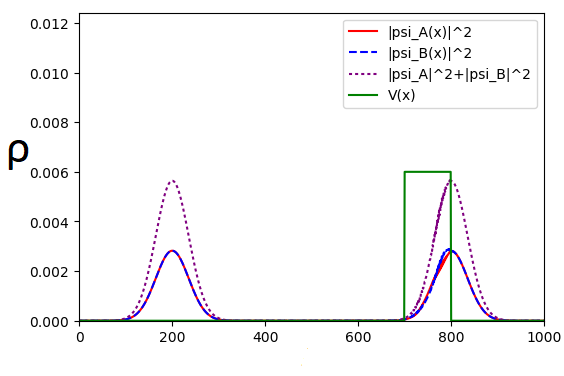} (b)&
			 \includegraphics[width=0.4\linewidth]{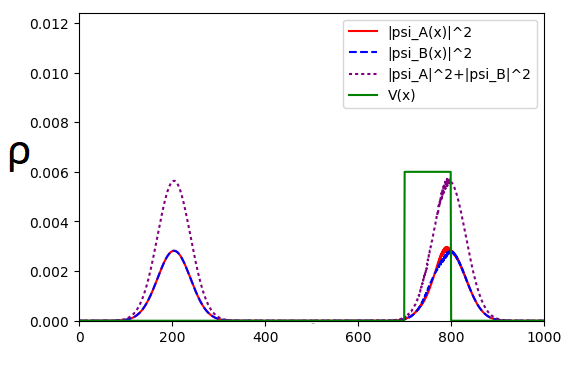} (e)\\
			 \includegraphics[width=0.4\linewidth]{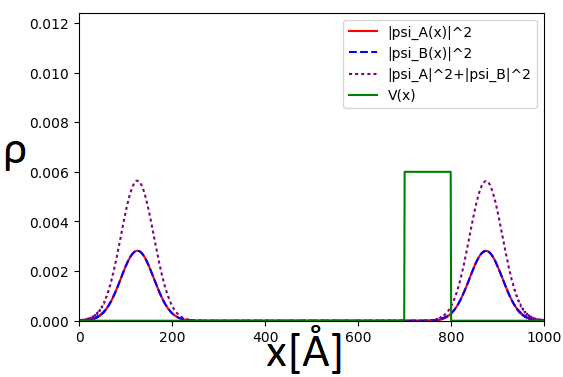}  (c)&
			 \includegraphics[width=0.4\linewidth]{fig2}   (f)\\
      \hline
\end{tabular}
\caption{\label{fig:comprash} Qualitative comparison of the collision with a QB of height $V_{b}=0.3$ eV, thickness $x_{b} = 100$ \AA, Rashba linear parameter $\alpha=0.4$ eV \AA $\,$ and Rashba cubic parameter $\beta=0.3$ eV\AA$^3$. We have taken $k_{x,y}=0$ (left-column panels) and $k_{x(y)}=0.09(0)$ \AA$^{-1}$ (right-column panels).}
\end{figure}

Figure \ref{fig:comprash} displays what happens with the 1DGWP after $t=12.5$ fs, considering the interaction with a QB under SOI-R. We can observe for $k = 0$ (left-column panels), that DF do not interplay with the SOI-R biased QB in any perceptible way. Meanwhile, when $k = 0.09$ \AA$^{-1}$ (right-column panels), one can see a weak splitting of the 1DGWP when it enters the QB, being $|\psi_{B}|^{2}$ slightly above [see blue-dashed line in panel (d)], otherwise when leaving the QB, is the component $|\psi_{A}|^{2}$ that clearly goes up [see red-solid line in panel (e)]. Even these differences, though quite small, between the cases without PZBE ($k =0$)[see left-column panels of Fig.\ref{fig:Zitcomp}], and with PZBE ($k =0.09$ \AA$^{-1}$)[see right-column panels of Fig.\ref{fig:Zitcomp}], the output 1DGWP is the same. The later has two possible interpretations: (i) From one side, we conclude that the coordinates of movement we have assumed are those of the normal incidence to the QB, then the Klein's paradox undergoes and the tunneling becomes perfect (transmission probability equals to one). (ii) On the other hand, it means that the QB and the SOI-R do not affect substantially the 1DGWP dynamics of the DF during their passage throughout the SOI-biased scatterer. Thus, the assessment of the SOI-R weakness in graphene\cite{Geimk2007}, is now reassured in the present study for the PZBE for the MLG configuration, because the way it becomes affected by the SOI-R is so far almost unnoticeable. Pursuing further confirmations, we have simulated the dynamics of an identical 1DGWP in MLG quantum wells. Although not shown here, due to possible numerical artifacts \textit{via} several finite-differences method inconsistences on the borders yet to be debugged, we have detected noticeable evidences of the PZBE modulated by the SOI-R. This may re-open the believe about the potential existence of an interplay between the ZBE and the SOI-R in graphene, following the sensible dependence of the ZBE from the SOI-R strength in $III-V$ semiconducting quantum wells\cite{Biswas2014}.

\begin{figure}
\begin{tabular}{|c|c|}
\hline
&\\
\includegraphics[width=0.4\linewidth]{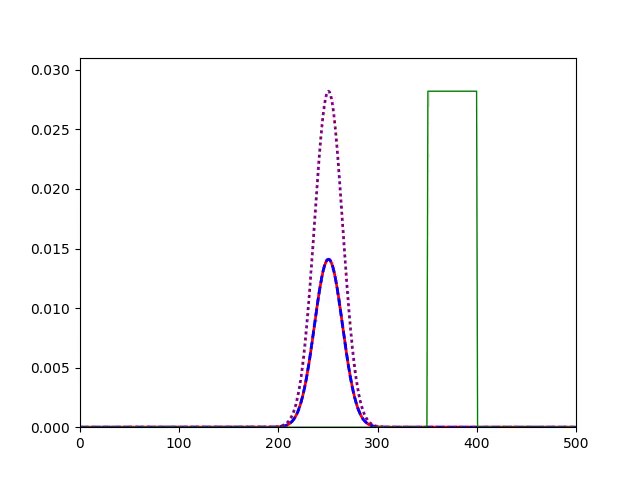}  (a)&
\includegraphics[width=0.4\linewidth]{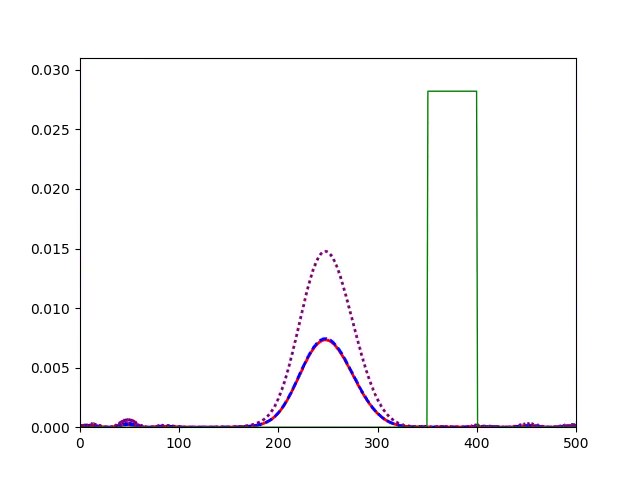}  (g)\\
\includegraphics[width=0.4\linewidth]{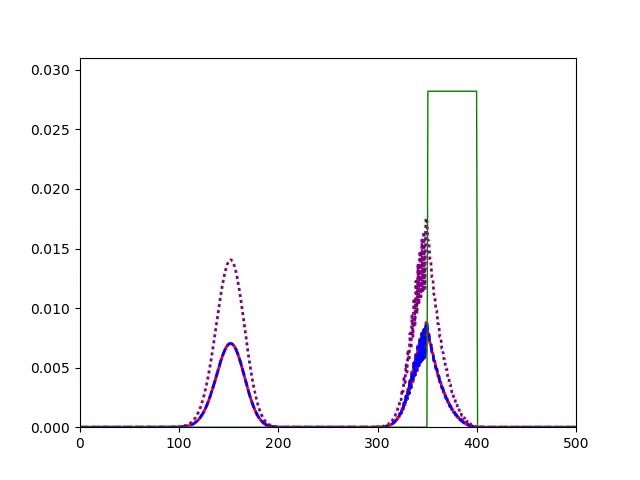}  (b)&
\includegraphics[width=0.4\linewidth]{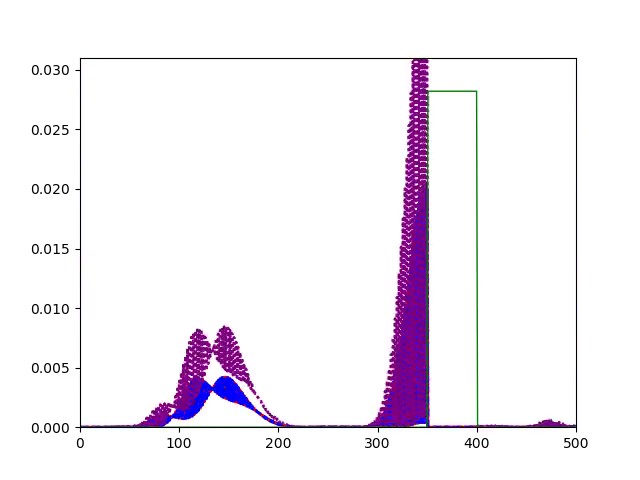}  (h)\\
\includegraphics[width=0.4\linewidth]{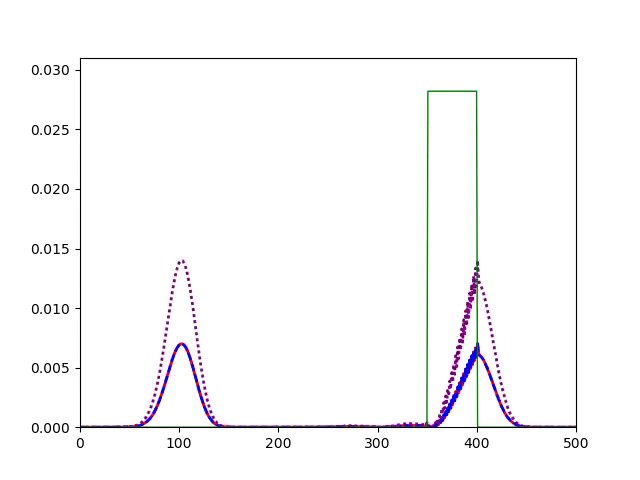}  (c)&
\includegraphics[width=0.4\linewidth]{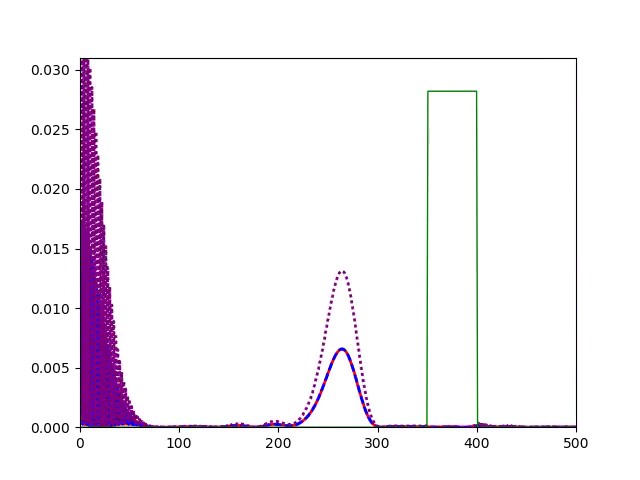}  (i)\\
\includegraphics[width=0.4\linewidth]{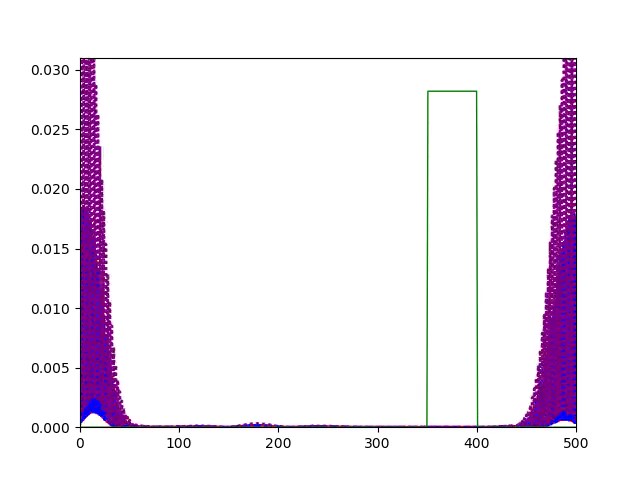}  (d)&
\includegraphics[width=0.4\linewidth]{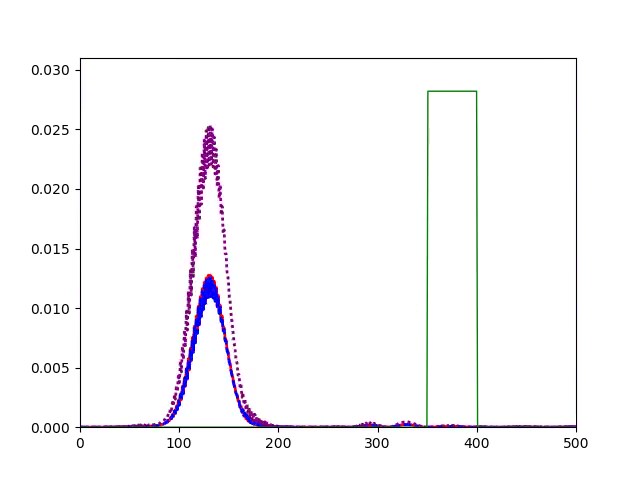}  (j)\\
\includegraphics[width=0.4\linewidth]{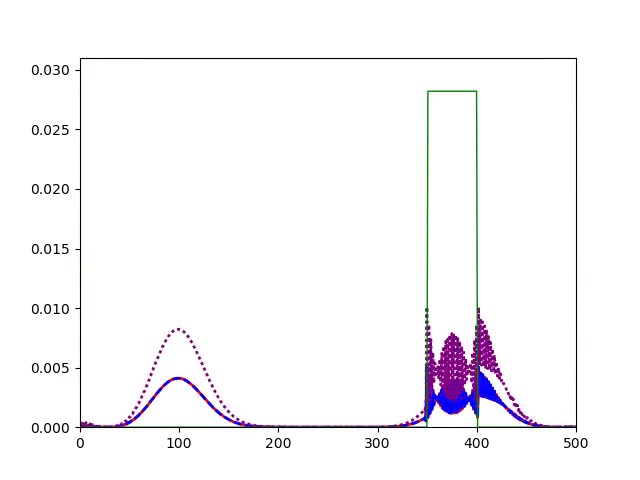}  (e)&
\includegraphics[width=0.4\linewidth]{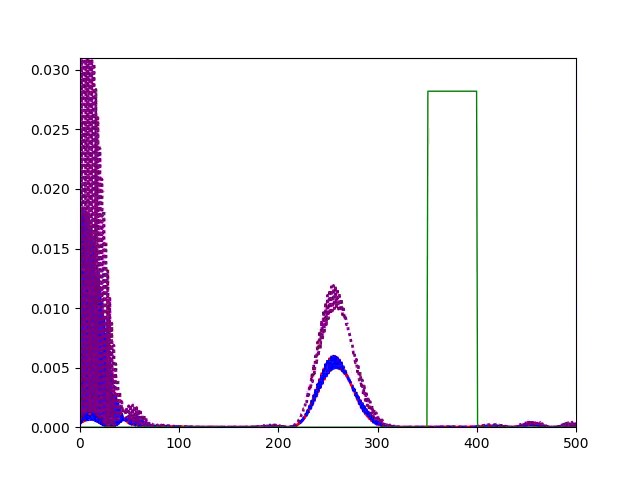}  (k)\\
\includegraphics[width=0.4\linewidth]{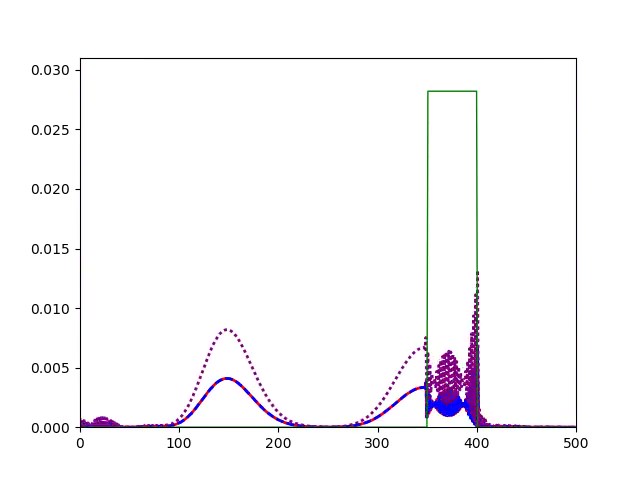}  (f)&
\includegraphics[width=0.4\linewidth]{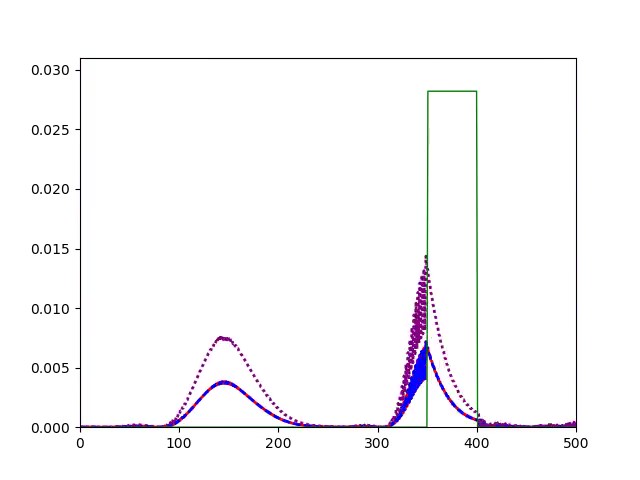}  (l)\\
\hline
\end{tabular}
\caption{\label{fig:totref} Complete temporal evolution of the WP with $\xi=\binom{1}{i}$ colliding with a barrier of height $V_b=0.3$ eV, width $x_b = 100$ \AA, Rashba linear parameter $\alpha=0.4$ eV \AA $\,$ and Rashba cubic parameter $\beta=0.3$ eV\AA$^3$; with $k=0$. We can see a complete transmission as well as a complete reflection.}
\end{figure}

Figure \ref{fig:totref}, for purpose of illustration, exposes an extended sequence of the 1DGWP time evolution inside a $100$ \AA$\,$ length surrounding box, to present evidences for a perfect backscattering of the 1DGWP at the QB under SOI-R [see animations of Ref.\onlinecite{AnimaQuantrix2018}]. On panels (a)-(d) a similar behavior as the one discussed in Fig.\ref{fig:comprash} can be seen and we have include them for the sake of completeness. Panel (e), plots the first interaction with the QB after the initial right-hand moving (RHM) sub-1DGWP, becomes reflected by the surrounding box wall. Notice some small portions of the sub-1DGWP due to multiple reflections with the QB and the box. They are traveling one way and the other way around, at the same time as they can be disregarded in most of this analysis given they somewhat represent a cavity noise. In panel (f), the backscattered RHM sub-1DGWP is trespassing the QB moving to the left, while the initial left-hand moving (LHM) sub-1DGWP evolves to the right. Panel (g), shows both sub-1DGWP overlapped and the noise allocated both sides of it. Next, on panel (h) there is an appealing total backscattering of the rearranged RHM sub-1DGWP with the QB. Meanwhile the LHM sub-1DGWP shape is being distorted by the cavity noise. Panel (i), exhibits the LHM sub-1DGWP collision with the box wall, when the perfect backscattered RHM sub-1DGWP is traveling to the left. On panel (j) they are overlapped one more time, now at the position $x\approx 120$ \AA. After that, panels (k) and (l) demonstrate a behavior quite similar to that of panels (a)-(d), except for the sub-1DGWP shape distortion, boosted by the cavity noise.

\section{\label{sec:Sum} Conclusions}
 Trembling anti-phase oscillations in the probability density time-distribution for each pseudo-spinor, which we have called pseudospinorial Zitterbewegung effect, are predicted as a new phenomenon of graphene. A novel nano-device is presented, that takes advantage of the peculiar morphologic characteristic of the $SiO_{2}$ substrate, the unique chemical properties of the $hBN$ and the gating techniques. The proposed setup, triggers SOI-R in all over the MLG sheet, generates tunable locally-induced SOI-R and preserves --in the same shoot--, the emblematic Dirac-cones shape for a gapless energy dispersion law. The theoretical model we have introduced, for describing the dynamics of DF in a Q1D-MLG, departs from the standardized Byschkov-Rashba Hamiltonian.  Good qualitative agreement with previous reports of the ZBE in graphene\cite{Frolova2008,Rusin2008,Rusin2011,Richter2012}, have been achieved, because a robust transient character of the PZBE oscillations, with decay time of about $10.5$ femtoseconds, was found in all cases under examination here. When sampling the vicinity of the $\Gamma$ and Dirac points $K$($K'$), we have observed that the frequency of the PZBE and the 1DGWP group velocity depend on the closeness to these high symmetry points. For example: when $k\rightarrow\Gamma$, the group velocity $\rightarrow 0$, while the frequency maximizes. Several features of the PZBE become tunable, (including it complete disappearance) as a function of $k$ and the initial pseudospin configuration. The SOI-R is very weak in MLG as reported before\cite{Geim2006}, and its influence on the PZBE was not yet detected. We have observed evidences of perfect Klein tunneling, together with perfect anti-Klein backscattering within the same simulation, which is quite unfamiliar for Q1D-MLG.

\section*{Acknowledgments}
We thanks Elizabeth Guti\'{e}rrez D\'{\i}az, Diablo Estudio Creativo, CDMX; for the design of Figs.\ref{fig:cells} and \ref{fig:modfis}. ES is indebted to  Salvador Carrillo, Departamento de F\'{\i}sica y Matem\'{a}ticas, Universidad Iberoamericana, CDMX, M\'{e}xico; for his teaching in parallel-computing techniques.

\bibliographystyle{apsrev4-1}
\bibliography{mybib6}

\begin{thebibliography}{20}%
\makeatletter
\providecommand \@ifxundefined [1]{%
 \@ifx{#1\undefined}
}%
\providecommand \@ifnum [1]{%
 \ifnum #1\expandafter \@firstoftwo
 \else \expandafter \@secondoftwo
 \fi
}%
\providecommand \@ifx [1]{%
 \ifx #1\expandafter \@firstoftwo
 \else \expandafter \@secondoftwo
 \fi
}%
\providecommand \natexlab [1]{#1}%
\providecommand \enquote  [1]{``#1''}%
\providecommand \bibnamefont  [1]{#1}%
\providecommand \bibfnamefont [1]{#1}%
\providecommand \citenamefont [1]{#1}%
\providecommand \href@noop [0]{\@secondoftwo}%
\providecommand \href [0]{\begingroup \@sanitize@url \@href}%
\providecommand \@href[1]{\@@startlink{#1}\@@href}%
\providecommand \@@href[1]{\endgroup#1\@@endlink}%
\providecommand \@sanitize@url [0]{\catcode `\\12\catcode `\$12\catcode
  `\&12\catcode `\#12\catcode `\^12\catcode `\_12\catcode `\%12\relax}%
\providecommand \@@startlink[1]{}%
\providecommand \@@endlink[0]{}%
\providecommand \url  [0]{\begingroup\@sanitize@url \@url }%
\providecommand \@url [1]{\endgroup\@href {#1}{\urlprefix }}%
\providecommand \urlprefix  [0]{URL }%
\providecommand \Eprint [0]{\href }%
\providecommand \doibase [0]{http://dx.doi.org/}%
\providecommand \selectlanguage [0]{\@gobble}%
\providecommand \bibinfo  [0]{\@secondoftwo}%
\providecommand \bibfield  [0]{\@secondoftwo}%
\providecommand \translation [1]{[#1]}%
\providecommand \BibitemOpen [0]{}%
\providecommand \bibitemStop [0]{}%
\providecommand \bibitemNoStop [0]{.\EOS\space}%
\providecommand \EOS [0]{\spacefactor3000\relax}%
\providecommand \BibitemShut  [1]{\csname bibitem#1\endcsname}%
\let\auto@bib@innerbib\@empty
%</preamble>
\bibitem [{\citenamefont {Ming-Hao~Liu}\ and\ \citenamefont
  {Richter}(2012)}]{Richter2012}%
  \BibitemOpen
  \bibfield  {author} {\bibinfo {author} {\bibfnamefont {J.~B.}\ \bibnamefont
  {Ming-Hao~Liu}}\ and\ \bibinfo {author} {\bibfnamefont {K.}~\bibnamefont
  {Richter}},\ }\href@noop {} {\bibfield  {journal} {\bibinfo  {journal} {Phys.
  Rev. B}\ }\textbf {\bibinfo {volume} {85}},\ \bibinfo {pages} {085406}
  (\bibinfo {year} {2012})}\BibitemShut {NoStop}%
\bibitem [{\citenamefont {Han}\ and\ \citenamefont
  {Fabian}(2014)}]{Fabian2014}%
  \BibitemOpen
  \bibfield  {author} {\bibinfo {author} {\bibfnamefont {W.}~\bibnamefont
  {Han}}\ and\ \bibinfo {author} {\bibfnamefont {J.}~\bibnamefont {Fabian}},\
  }\href@noop {} {\bibfield  {journal} {\bibinfo  {journal} {Nat. Nanotech.}\
  }\textbf {\bibinfo {volume} {9}},\ \bibinfo {pages} {794} (\bibinfo {year}
  {2014})}\BibitemShut {NoStop}%
\bibitem [{\citenamefont {Geim}\ and\ \citenamefont
  {Novoselov}(2007)}]{Geimk2007}%
  \BibitemOpen
  \bibfield  {author} {\bibinfo {author} {\bibfnamefont {A.~K.}\ \bibnamefont
  {Geim}}\ and\ \bibinfo {author} {\bibfnamefont {K.~S.}\ \bibnamefont
  {Novoselov}},\ }\href@noop {} {\bibfield  {journal} {\bibinfo  {journal}
  {Nat. Mater.}\ }\textbf {\bibinfo {volume} {6}},\ \bibinfo {pages} {183}
  (\bibinfo {year} {2007})}\BibitemShut {NoStop}%
\bibitem [{\citenamefont {Neto}(2009)}]{Castro2009}%
  \BibitemOpen
  \bibfield  {author} {\bibinfo {author} {\bibfnamefont {A.~H.~C.}\
  \bibnamefont {Neto}},\ }\href@noop {} {\bibfield  {journal} {\bibinfo
  {journal} {Rev. Mod. Phys}\ }\textbf {\bibinfo {volume} {81}},\ \bibinfo
  {pages} {109} (\bibinfo {year} {2009})}\BibitemShut {NoStop}%
\bibitem [{\citenamefont {Rodriguez}\ and\ \citenamefont
  {Vasilievna}(2008)}]{Vasilievna2008}%
  \BibitemOpen
  \bibfield  {author} {\bibinfo {author} {\bibfnamefont {C.}~\bibnamefont
  {Rodriguez}}\ and\ \bibinfo {author} {\bibfnamefont {O.}~\bibnamefont
  {Vasilievna}},\ }\href@noop {} {\bibfield  {journal} {\bibinfo  {journal}
  {Ingenierias, UANL, (ISSN: \textbf{1405-0676}, http://ingenierias.uanl.mx/)}\
  }\textbf {\bibinfo {volume} {11}},\ \bibinfo {pages} {17} (\bibinfo {year}
  {2008})}\BibitemShut {NoStop}%
\bibitem [{\citenamefont {Schr\"{o}dinger}(1930)}]{Schro1930}%
  \BibitemOpen
  \bibfield  {author} {\bibinfo {author} {\bibfnamefont {E.}~\bibnamefont
  {Schr\"{o}dinger}},\ }\href@noop {} {\bibfield  {journal} {\bibinfo
  {journal} {Sitzungsb. Preuss. Akad. Wiss. Phys.-Math. Kl.}\ }\textbf
  {\bibinfo {volume} {24}},\ \bibinfo {pages} {418} (\bibinfo {year}
  {1930})}\BibitemShut {NoStop}%
\bibitem [{\citenamefont {Dirac}(1958)}]{Dirac1958}%
  \BibitemOpen
  \bibfield  {author} {\bibinfo {author} {\bibfnamefont {P.}~\bibnamefont
  {Dirac}},\ }\href@noop {} {\emph {\bibinfo {title} {The Principles of Quantum
  Mechanics}}}\ (\bibinfo  {publisher} {Clarendon Press, Oxford},\ \bibinfo
  {year} {1958})\ pp.\ \bibinfo {pages} {4ff, 253ff}\BibitemShut {NoStop}%
\bibitem [{\citenamefont {Maksimova}\ and\ \citenamefont
  {Frolova}(2008)}]{Frolova2008}%
  \BibitemOpen
  \bibfield  {author} {\bibinfo {author} {\bibfnamefont {G.~M.}\ \bibnamefont
  {Maksimova}}\ and\ \bibinfo {author} {\bibfnamefont {E.~V.}\ \bibnamefont
  {Frolova}},\ }\href@noop {} {\bibfield  {journal} {\bibinfo  {journal} {Phys.
  Rev. B}\ }\textbf {\bibinfo {volume} {78}},\ \bibinfo {pages} {235321}
  (\bibinfo {year} {2008})}\BibitemShut {NoStop}%
\bibitem [{\citenamefont {Biswas}\ and\ \citenamefont
  {Ghosh}(2014)}]{Biswas2014}%
  \BibitemOpen
  \bibfield  {author} {\bibinfo {author} {\bibfnamefont {T.}~\bibnamefont
  {Biswas}}\ and\ \bibinfo {author} {\bibfnamefont {T.~K.}\ \bibnamefont
  {Ghosh}},\ }\href@noop {} {\bibfield  {journal} {\bibinfo  {journal} {J.
  Appl. Phys.}\ }\textbf {\bibinfo {volume} {115}} (\bibinfo {year}
  {2014})}\BibitemShut {NoStop}%
\bibitem [{\citenamefont {P.~Krekora}\ and\ \citenamefont
  {Grobe}(2004)}]{Krekora2004}%
  \BibitemOpen
  \bibfield  {author} {\bibinfo {author} {\bibfnamefont {Q.~S.}\ \bibnamefont
  {P.~Krekora}}\ and\ \bibinfo {author} {\bibfnamefont {R.}~\bibnamefont
  {Grobe}},\ }\href@noop {} {\bibfield  {journal} {\bibinfo  {journal} {Phys.
  Rev. Lett.}\ }\textbf {\bibinfo {volume} {93}},\ \bibinfo {pages} {043004}
  (\bibinfo {year} {2004})}\BibitemShut {NoStop}%
\bibitem [{\citenamefont {Katsnelson}(2006)}]{Katsnelson2006}%
  \BibitemOpen
  \bibfield  {author} {\bibinfo {author} {\bibfnamefont {M.~I.}\ \bibnamefont
  {Katsnelson}},\ }\href@noop {} {\bibfield  {journal} {\bibinfo  {journal}
  {Eur. Phys. J. B}\ }\textbf {\bibinfo {volume} {51}},\ \bibinfo {pages} {157}
  (\bibinfo {year} {2006})}\BibitemShut {NoStop}%
\bibitem [{\citenamefont {Sidhart}(2008)}]{Sidhart2008}%
  \BibitemOpen
  \bibfield  {author} {\bibinfo {author} {\bibfnamefont {B.~G.}\ \bibnamefont
  {Sidhart}},\ }\href@noop {} {\bibfield  {journal} {\bibinfo  {journal}
  {arXiv:0806.0985v1[physics-gen-ph]}\ } (\bibinfo {year} {2008})}\BibitemShut
  {NoStop}%
\bibitem [{\citenamefont {Rusin}\ and\ \citenamefont
  {Zawadski}(2008)}]{Rusin2008}%
  \BibitemOpen
  \bibfield  {author} {\bibinfo {author} {\bibfnamefont {T.}~\bibnamefont
  {Rusin}}\ and\ \bibinfo {author} {\bibfnamefont {W.}~\bibnamefont
  {Zawadski}},\ }\href@noop {} {\bibfield  {journal} {\bibinfo  {journal}
  {Phys. Rev. B}\ }\textbf {\bibinfo {volume} {78}},\ \bibinfo {pages} {125419}
  (\bibinfo {year} {2008})}\BibitemShut {NoStop}%
\bibitem [{\citenamefont {Zawadski}\ and\ \citenamefont
  {Rusin}(2011)}]{Rusin2011}%
  \BibitemOpen
  \bibfield  {author} {\bibinfo {author} {\bibfnamefont {W.}~\bibnamefont
  {Zawadski}}\ and\ \bibinfo {author} {\bibfnamefont {T.}~\bibnamefont
  {Rusin}},\ }\href@noop {} {\bibfield  {journal} {\bibinfo  {journal}
  {arXiv:1101.0623v1[cond-mat-mes-hall]}\ } (\bibinfo {year}
  {2011})}\BibitemShut {NoStop}%
\bibitem [{\citenamefont {Min}\ and\ \citenamefont
  {MacDonald}(2006)}]{MacDonald2006}%
  \BibitemOpen
  \bibfield  {author} {\bibinfo {author} {\bibfnamefont {H.}~\bibnamefont
  {Min}}\ and\ \bibinfo {author} {\bibfnamefont {A.~H.}\ \bibnamefont
  {MacDonald}},\ }\href@noop {} {\bibfield  {journal} {\bibinfo  {journal}
  {Phys. Rev. B}\ }\textbf {\bibinfo {volume} {74}},\ \bibinfo {pages} {165310}
  (\bibinfo {year} {2006})}\BibitemShut {NoStop}%
\bibitem [{\citenamefont {Cuan}\ and\ \citenamefont
  {Diago-Cisneros}(2010)}]{RDiago2010}%
  \BibitemOpen
  \bibfield  {author} {\bibinfo {author} {\bibfnamefont {R.}~\bibnamefont
  {Cuan}}\ and\ \bibinfo {author} {\bibfnamefont {L.}~\bibnamefont
  {Diago-Cisneros}},\ }\href@noop {} {\bibfield  {journal} {\bibinfo  {journal}
  {Rev. Cub. F\'{\i}s.}\ }\textbf {\bibinfo {volume} {27}},\ \bibinfo {pages}
  {212} (\bibinfo {year} {2010})}\BibitemShut {NoStop}%
\bibitem [{\citenamefont {Cuan}\ and\ \citenamefont
  {Diago-Cisneros}(2015)}]{RCDiagoEPL2015}%
  \BibitemOpen
  \bibfield  {author} {\bibinfo {author} {\bibfnamefont {R.}~\bibnamefont
  {Cuan}}\ and\ \bibinfo {author} {\bibfnamefont {L.}~\bibnamefont
  {Diago-Cisneros}},\ }\href@noop {} {\bibfield  {journal} {\bibinfo  {journal}
  {Eur. Phys. Lett.}\ }\textbf {\bibinfo {volume} {110}},\ \bibinfo {pages}
  {67001} (\bibinfo {year} {2015})}\BibitemShut {NoStop}%
\bibitem [{Ani(2018)}]{AnimaQuantrix2018}%
  \BibitemOpen
  \href@noop {} {\bibfield  {journal} {\bibinfo  {journal}
  {https://sernaed95.wixsite.com/animaquantrix/animations \\Multimedia of the
  1DGWP wordline dynamic}\ } (\bibinfo {year} {2018})}\BibitemShut {NoStop}%
\bibitem [{\citenamefont {Carrillo}\ and\ \citenamefont
  {Mendoza}(2015)}]{Carrillo2015}%
  \BibitemOpen
  \bibfield  {author} {\bibinfo {author} {\bibfnamefont {A.}~\bibnamefont
  {Carrillo}}\ and\ \bibinfo {author} {\bibfnamefont {O.}~\bibnamefont
  {Mendoza}},\ }\href@noop {} {\bibfield  {journal} {\bibinfo  {journal}
  {Geofisica UNAM}\ } (\bibinfo {year} {2015})}\BibitemShut {NoStop}%
\bibitem [{\citenamefont {Katsnelson}\ and\ \citenamefont
  {Geim}(2006)}]{Geim2006}%
  \BibitemOpen
  \bibfield  {author} {\bibinfo {author} {\bibfnamefont {M.~I.}\ \bibnamefont
  {Katsnelson}}\ and\ \bibinfo {author} {\bibfnamefont {A.~K.}\ \bibnamefont
  {Geim}},\ }\href@noop {} {\bibfield  {journal} {\bibinfo  {journal} {Nat.
  Phys.}\ }\textbf {\bibinfo {volume} {2}},\ \bibinfo {pages} {620} (\bibinfo
  {year} {2006})}\BibitemShut {NoStop}%
\end{thebibliography}%
\newpage

%\nocite{*}
\end{document}